\documentclass[journal]{IEEEtran}
\ifCLASSINFOpdf
\else
\fi
\usepackage{amsmath}
\usepackage{amsfonts}
\usepackage{booktabs}
\usepackage{multirow}
\usepackage{bm}
\usepackage{booktabs,caption}
\usepackage[flushleft]{threeparttable}
\usepackage{amsmath,graphicx}
\usepackage{enumitem}
\usepackage{amsthm}
\usepackage{amssymb}
\usepackage{mathtools}
\usepackage{xcolor} 
\usepackage{ulem} 
\RequirePackage[colorlinks]{hyperref}
\newcommand{\widesim}[2][1.5]{
  \mathrel{\overset{#2}{\scalebox{#1}[1]{$\sim$}}}
}

\DeclareMathOperator{\Var}{Var}

\newtheorem{thm}{Theorem}[section]
\newtheorem{Lemma}{Lemma}

\definecolor{bcol}{rgb}{0.8,0,0.2}

\begin{document}
%
\title{Estimation of the Hurst Exponent
\\ Using Trimean Estimators on \\ Nondecimated Wavelet Coefficients}
%
%
%

\author{Chen~Feng~and~Brani~Vidakovic
\thanks{}
\thanks{}
\thanks{}}

%
%

\markboth{Journal of \LaTeX\ Class Files,~Vol.~14, No.~8, August~2015}%
{Shell \MakeLowercase{\textit{et al.}}: Bare Demo of IEEEtran.cls for IEEE Journals}
%



\maketitle

\begin{abstract}
Hurst exponent 
is an important feature summarizing the noisy high-frequency data when the inherent scaling pattern cannot be described by standard statistical models. In this paper, we study the robust estimation of Hurst exponent based on non-decimated wavelet transforms (NDWT). The robustness is achieved by applying a general trimean estimator on non-decimated wavelet coefficients of the transformed data. The general trimean estimator is derived as a weighted average of the distribution's median and quantiles, combining the median's emphasis on central values with the quantiles' attention to the extremes. The properties of the proposed Hurst exponent estimators  are studied both theoretically and numerically. Compared with other standard wavelet-based methods (Veitch $\&$ Abry (VA) method, Soltani, Simard, $\&$ Boichu (SSB) method, median based estimators MEDL and MEDLA), our methods reduce the variance of the estimators and increase the prediction precision in most cases. The proposed methods are applied to a data set in high frequency pupillary response behavior (PRB) with the goal to classify individuals according to a degree of their visual impairment.
\end{abstract}


%
\IEEEpeerreviewmaketitle

\section{Introduction}
%
%
%
%
\IEEEPARstart{H}{igh-frequency}, time series data from various sources often possess hidden patterns that reveal the effects of underlying functional differences, but such patterns cannot be explained by basic descriptive statistics, traditional statistical models, or global trends. For example, the high-frequency pupillary response behavior (PRB) data collected during human-computer interaction  capture  the changes in pupil diameter in response to simple or complex stimuli. Researchers found that there may be underlying unique patterns hidden within complex PRB data, and these patterns reveal the intrinsic individual differences in cognitive, sensory and motor functions \cite{Moloney:2006:LDC:1183456.1183460}. Yet, such patterns cannot be explained by the traditional statistical summaries of the PRB data, for the magnitude of the diameter depends on ambient light, not on the inherent eye function \cite{2017arXiv170304180K}. When the intrinsic individual functional differences cannot be described by basic statistics and trends in the noisy high-frequency data, the Hurst exponent might be an optional measure of patients' characterization.

The Hurst exponent $H$ quantifies the long memory as well as regularity, self-similarity, and scaling in a time series. Among models having been proposed for analyzing the self-similar phenomena, arguably the most popular is the fractional Brownian motion (fBm) first described by Kolmogorov \cite{Kolmogorov1940} and formalized by Mandelbrot and Van Ness \cite{mandelbrot1968fractional}. Its importance is due to the fact that fBm is the unique Gaussian process with stationary increments that is self-similar. Recall that a stochastic process $\left\{X\left(\bm{t}\right), \bm{t}\in \mathbb{R}^d\right\}$ is self-similar with Hurst exponent $H$ if, for any $\lambda\in\mathbb{R}^+$, $X\left(\bm{t}\right)\overset{\mathrm{d}}{=}\lambda^{-H}X\left(\lambda\bm{t}\right)$. Here the notation $\overset{\mathrm{d}}{=}$ means the equality in all finite-dimensional distributions. Hurst exponent $H$ describes the rate at which autocorrelations decrease as the lag between two realizations in a time series increases.

A value $H$ in the range 0-0.5 indicates a zig-zagging intermittent time series with long-term switching between high and low values in adjacent pairs. 
A value $H$ in the range 0.5 to 1 indicates a time series with long-term positive autocorrelations, which preserves trends on a longer time horizon and gives a time series more regular appearance.


Multiresolution analysis is one of the many methods to estimate the Hurst exponent $H$. An overview can be found in \cite{abry2000wavelets,abry1995wavelets,abry2013scaling}. If processes possess a stochastic structure, i.e., Gaussianity and stationary increments, $H$ becomes a parameter in a well-defined statistical model and can be estimated after taking the $\log_2$ of the quantity $\mathbb{E}\left(d_j^2\right)$, where $d_j$'s are the wavelet coefficients at level $j$. In fact, several estimation methods of $H$ based on wavelets analysis exist. Veitch and Abry \cite{abry2000wavelets} suggest the estimation of $H$ by  weighted least square regression using the level-wise $\log_2\left(\bar{d_j^2}\right)$ since the variance of $\log_2\left(\bar{d_j^2}\right)$ depends on $H$ and the level $j$. In addition, the authors correct for the bias caused by the order of taking the logarithm and the average in $\log_2\left(\bar{d_j^2}\right)$. Soltani et al \cite{soltani2004estimation} defined a mid-energy as $D_{j,k}=\left(d_{j,k}^2+d_{j,k+N_j/2}^2\right)/2$, and showed that the level-wise averages of $\log_2D_{j,k}$ are asymptotically normal and more stable, which is used to estimate parameter $H$ by regression. The estimators in Soltani et al \cite{soltani2004estimation} consistently outperform the estimators in Veitch and Abry \cite{abry2000wavelets}. Shen et al \cite{shen2007robust} shows that the method of Soltani et al \cite{soltani2004estimation} yields more accurate estimators since it takes the logarithm of the mid-energy and then averages. Kang and Vidakovic \cite{2017arXiv170304180K} proposed MEDL and MEDLA methods based on non-decimated wavelets to estimate $H$. MEDL estimates $H$ by regressing the medians of $\log d_j^2$ on level $j$, while MEDLA uses the level-wise medians of $\log\left(\left(d_{j,k_1}^2+d_{j,k_2}^2\right)/2\right)$ to estimate $H$, where $k_1$ and $k_2$ are properly selected locations at level $j$ to guarantee the independence assumption. Both MEDL and MEDLA use the median of the derived distribution instead of the mean, because the medians are more robust to potential outliers that can occur when logarithmic transform of a squared wavelet coefficient is taken and the magnitude of coefficient is close to zero. Although median is outlier-resistant, it can behave unexpectedly as a result of its non-smooth character. The fact that the median is not ``universally the best outlier-resistant estimator" provides a practical motivation for examining alternatives that are intermediate in behavior between the very smooth but outlier-sensitive mean and the very outlier-insensitive but non-smooth median. In this article, the general trimean estimator is derived as a weighted average of the distribution's median and two quantiles symmetric about the median, combining the median's emphasis on center values with the quantiles' attention to the tails. Tukey's trimean estimator \cite{tukey1977exploratory, andrews2015robust} and Gastwirth estimator \cite{Gas1, Gas2, Gas3} turn out to be two special cases under the general framework. We will use the general trimean estimators of the level-wise derived distributions to estimate $H$. In this paper, we are concerned with the robust estimation of Hurst exponent in one-dimensional setting, however, the methodology can be readily extended to a multidimensional case.

The rest of the paper consists of five additional sections and an appendix. Section 2 introduces the general trimean estimators and discusses two special estimators following that general framework; Section 3 describes estimation of Hurst exponent using the general trimean estimators, presents distributional results on which the proposed methods are based, and derives  optimal weights that minimize the  variances of the estimators. Section 4 provides the simulation results and compares the  performance of the proposed methods to  other standardly used, wavelet-based methods. The proposed methods are illustrated using the real PRB data in Section 5. The paper is concluded with a summary and discussion in Section 6.

\section{General Trimean Estimators}\label{section general}
Let $X_1, X_2,..., X_n$ be i.i.d. continuous random variables with pdf $f(x)$ and cdf $F(x)$. Let $0<p<1$, and let $\xi_p$ denote the $p$th quantile of $F$, so that
$\xi_p = \inf\{x| F(x) \geq p\}.$ If $F$ is monotone, the $p$th quantile is simply defined as
$F(\xi_p)=p$.

Let $Y_p=X_{\lfloor np\rfloor:n}$ denote a sample $p$th quantile. Here $\lfloor np\rfloor$ denotes the greatest integer that is less than or equal to $np$. The general trimean estimator is defined as a weighted average of the distribution's median and its two quantiles $Y_{p}$ and $Y_{1-p}$, for $p\in(0,1/2)$:
\begin{equation}\label{eq:gte1}
\hat{\mu}=\frac{\alpha}{2}\ Y_{p}+\left(1-\alpha\right)\ Y_{1/2}+\frac{\alpha}{2}\ Y_{1-p}.
\end{equation}
The weights for the two quantiles are the same for $Y_{p}$ and $Y_{1-p}$, and $\alpha\in[0,1]$. This is equivalent to the weighted sum of the median and the average of $Y_{p}$ and $Y_{1-p}$ with weights $1-\alpha$ and $\alpha$:
\begin{equation*}
\hat{\mu}=\left(1-\alpha\right)\ Y_{1/2}+\alpha\ \left(\frac{Y_{p}+Y_{1-p}}{2}\right).
\end{equation*}
This general trimean estimator turns out to be more robust then mean but smoother than the median. To derive the asymptotic distribution of this general trimean estimator, the asymptotic joint distribution of sample quantiles is shown in Lemma 1; detailed proof can be found in \cite{dasgupta2008asymptotic}.

\begin{Lemma}\label{Lemma:L1}
Consider $r$ sample quantiles, $Y_{p_1}, Y_{p_2},....,Y_{p_r}$, where $1 \leq p_1<p_2<...<p_r\leq n$. If for any $1\le i\le r$, $\sqrt{n}\left(\lfloor np_i\rfloor/n-p_i\right)\to 0$ is satisfied, then the asymptotic joint
 distribution of $Y_{p_1}, Y_{p_2},....,Y_{p_r}$ is:
\[\sqrt{n}\left(\left(Y_{p_1}, Y_{p_2},....,Y_{p_r}\right)-\left(\xi_{p_1}, \xi_{p_2},....,\xi_{p_r}\right)\right)\widesim{\footnotesize{\mbox{approx}}} {\cal MVN}\left(0, \Sigma\right),\]
where
$$\Sigma=\left(\sigma_{ij}\right)_{r\times r},$$ and
\begin{equation}\label{eq:sigma}
\sigma_{ij}=\frac{p_i\left(1-p_j\right)}{f\left(x_{p_i}\right)f\left(x_{p_j}\right)},\ i\le j.
\end{equation}
\end{Lemma}

From Lemma $\ref{Lemma:L1}$, the asymptotic distribution of general trimean estimator will be normal as a linear combination of the components of the asymptotic multivariate normal distribution. The general trimean estimator itself may be defined in terms of order statistics as
\[\hat{\mu}=A\cdot\bm{y},\]
where
\[A=\left[\frac{\alpha}{2}\ \ \ 1-\alpha\ \ \ \frac{\alpha}{2}\right],\ \  \mbox{and}\ \ \bm{y}=\left[Y_{p}\ \ \ Y_{1/2}\ \ \ Y_{1-p}\right]^T.\]
It can be easily verified that $\sqrt{n}\left(\lfloor pn\rfloor/n-p\right)\to 0$ for $p\in\left(0,1/2\right]$. If we denote $\bm{\xi}=\left[\xi_{p}\ \ \ \xi_{1/2}\ \ \ \xi_{1-p}\right]^T$ the population quantiles, the asymptotic distribution of $\bm{y}$ is
\[\sqrt{n}\left(\bm{y}-\bm{\xi}\right)\widesim{\footnotesize{\mbox{approx}}} {\cal MVN}\left(0, \Sigma\right),\]
where $\Sigma=\left(\sigma_{ij}\right)_{3\times 3},$ and $\sigma_{ij}$ follows equation ($\ref{eq:sigma}$) for $p_1=p,\ p_2=1/2,\ \mbox{and}\ p_3=1-p$. Therefore
\[\hat{\mu}\widesim{\footnotesize{\mbox{approx}}} \mathcal{N}\left(\mathbb{E}\left(\hat{\mu}\right), \Var\left(\hat{\mu}\right)\right),\]
with the theoretical expectation and variance being
\begin{equation}
\mathbb{E}\left(\hat{\mu}\right)=\mathbb{E}\left(A\cdot\bm{y}\right)=A\cdot\mathbb{E}\left(\bm{y}\right)=A\cdot\bm{\xi},
\end{equation}
and
\begin{equation}
\Var\left(\hat{\mu}\right)=\Var\left(A\cdot\bm{y}\right)=A\Var\left(\bm{y}\right)A^T=\frac{1}{n}A\Sigma A^T.
\end{equation}


\subsection{Tukey's Trimean Estimator}\label{section Tukey}
Tukey's trimean estimator is a special case of the general trimean estimators, with $\alpha=1/2$ and $p=1/4$ in equation ($\ref{eq:gte1}$). To compute this estimator, we first sort the data in ascending order. Next, we take the values that are one-forth of the way up this sequence (the first quartile), half way up the sequence (i.e., the median),
and three-fourths of the way up the sequence (the thirt quartile).  Given these three values, we then form the weighted average, giving the central (median) value a weight of $1/2$ and the two quartiles each a weight of $1/4$. If we denote  the Tukey's trimean estimator
as $\hat{\mu}_T,$
 then
\[\hat{\mu}_T=\frac{1}{4}\ Y_{1/4}+\frac{1}{2}\ Y_{1/2}+\frac{1}{4}\ Y_{3/4}.\]
The Tukey's trimean estimator may be expressed more compactly as
\[\hat{\mu}_T=A_T\cdot\bm{y}_T,\]
where
\[A_T=\left[\frac{1}{4}\ \ \ \frac{1}{2}\ \ \ \frac{1}{4}\right],\ \  \mbox{and}\ \ \bm{y}_T=\left[Y_{1/4}\ \ \ Y_{1/2}\ \ \ Y_{3/4}\right]^T.\]
It can be easily verified that $\sqrt{n}\left(\lfloor pn\rfloor/n-p\right)\to 0$ for $p=1/4, 1/2, \mbox{and}\  3/4$. If we denote $\bm{\xi}_T=\left[\xi_{1/4}\ \ \ \xi_{1/2}\ \ \ \xi_{3/4}\right]^T$  the corresponding theoretical quantiles, the asymptotic distribution of $\hat{\mu}_T$ is
\[\hat{\mu}_T\widesim{\footnotesize{\mbox{approx}}} \mathcal{N}\left(A_T\cdot\bm{\xi}_T, \frac{1}{n}A_T\Sigma_T A_T^T\right),\]
where $\Sigma_T=\left(\sigma_{ij}\right)_{3\times 3}$ is the covariance matrix of the asymptotic multivariate normal distribution, and $\sigma_{ij}$ follows equation ($\ref{eq:sigma}$) with $p_1=1/4,\ p_2=1/2,\ \mbox{and}\ p_3=3/4$.

\subsection{Gastwirth Estimator}\label{section Gastwirth}
As the Tukey's estimator,  the Gastwirth estimator is another special case of the general trimean estimators, with $\alpha=0.6$ and $p=1/3$ in equation ($\ref{eq:gte1}$).
If we denote this estimator as $\hat{\mu}_G$, then
\[\hat{\mu}_G=0.3\ Y_{1/3}+0.4\ Y_{1/2}+0.3\ Y_{2/3}.\]
The Gastwirth estimator can be written as
\[\hat{\mu}_G=A_G\cdot\bm{t}_G,\]
where
\[A_G=\left[0.3\ \ \ 0.4\ \ \ 0.3\right],\ \  \mbox{and}\ \ \bm{y}_G=\left[Y_{1/3}\ \ \ Y_{1/2}\ \ \ Y_{2/3}\right]^T.\] As in Tukey's case, if we denote $\bm{\xi}_G=\left[\xi_{1/3}\ \ \ \xi_{1/2}\ \ \ \xi_{2/3}\right]^T$ as the theoretical quantiles, since $\sqrt{n}\left(\lfloor pn\rfloor/n-p\right)\to 0$ for $p=1/3, 1/2, \mbox{and}\  2/3$, the asymptotic distribution of $\hat{\mu}_G$ is
\[\hat{\mu}_G\widesim{\footnotesize{\mbox{approx}}} \mathcal{N}\left(A_G\cdot\bm{\xi}_G, \frac{1}{n}A_G\Sigma_G A_G^T\right),\]
where $\Sigma_G=\left(\sigma_{ij}\right)_{3\times 3}$, and $\sigma_{ij}$ follows equation ($\ref{eq:sigma}$) with $p_1=1/3,\ p_2=1/2,\ \mbox{and}\ p_3=2/3$.

\section{Methods}
Our proposal for robust estimation of Hurst exponent $H$ is based on non-decimated wavelet transforms (NDWT). Details of NDWT can be found in \cite{nason1995stationary, vidakovic2009statistical, percival2006wavelet}. The comparison of NDWT and standard orthogonal discrete wavelet transforms (DWT) has been discussed by Kang and Vidakovic \cite{2017arXiv170304180K}. NDWT has several advantages when employed for Hurst exponent estimation: 1) Input signals and images of arbitrary size can be processed due to the absence of decimation; 2) as a redundant transform, the NDWT increases the accuracy of the scaling estimation; 3) least square regression can be fitted to estimate $H$ instead of weighted least square regression since the variances of the level-wise derived distributions based on logged NDWT coefficients do not depend on level; 4) local scaling can be assessed due to the time-invariance property. Of course, as we will discuss later, the dependence of coefficients in NDWT is
much more pronounced than in the case of DWT.

In $J$-level decomposition of a fBm of size $N$, a NDWT yields $N\times \left(J+1\right)$ wavelet coefficients, with each level $N$ coefficients. At each level $j$, we generate $N/2$  mid-energies as
\begin{equation}
D_{j,k}=\left(d_{j,k}^2+d_{j,k+N/2}^2\right)/2,\ \mbox{for} \ k=1,2,...,N/2.
\end{equation}
The distributions of the $D_{j,k}$ and $\log{D_{j,k}}$ are derived under the assumption that $d_{j,k}$ and $d_{j,k+N/2}$ are independent. Then we apply the general trimean estimators of the level-wise derived distributions to estimate parameter $H$. Note that for fixed $j$, the generated $N/2$  mid-energies $D_{j,k}$ and $\log{D_{j,k}}$ are not independent but their autocorrelations decay exponentially, consequently, they posses only the short memory. At each level $j$, we sample every $M$ points from $D_{j,k}$ and $\log{D_{j,k}}$ to form $M$ groups, and assume that the $\left(N/2\right)/M$ points of $D_{j,k}$ and $\log{D_{j,k}}$ within each subgroup are independent, respectively. The general trimean estimators are then applied on each of the $M$ groups. Note that $M$ must be divisible by $N/2$.

\begin{tabular}{p{13cm}}
\small
\begin{description}
\item[Group 1:] $\left\{D_{j,1},D_{j,1+M},D_{j,1+2M},...,D_{j,\left(N/2-M+1\right)}\right\}$ and
\item[]\ \ \ \ \ \ \ \ \ \ \ \ $\left\{\log\left(D_{j,1}\right),\log\left(D_{j,1+M}\right),...,\log\left(D_{j,\left(N/2-M+1\right)}\right)\right\}$
\item[Group 2:] $\left\{D_{j,2},D_{j,2+M},D_{j,2+2M},...,D_{j,\left(N/2-M+2\right)}\right\}$ and
\item[]\ \ \ \ \ \ \ \ \ \ \ \ $\left\{\log\left(D_{j,2}\right),\log\left(D_{j,2+M}\right),...,\log\left(D_{j,\left(N/2-M+2\right)}\right)\right\}$
\item[] \vdots
\item[Group M:] $\left\{D_{j,M},D_{j,2M},D_{j,3M},...,D_{j,N/2}\right\}$ and
\item[]\ \ \ \ \ \ \ \ \ \ \ \ \ $\left\{\log\left(D_{j,M}\right),\log\left(D_{j,2M}\right),...,\log\left(D_{j,N/2}\right)\right\}$
\end{description}\\
\end{tabular}

\subsection{General Trimean of the Mid-energy  (GTME) Method}
For the general trimean of the  mid-energy  (GTME) method, we derive the relationship between the general trimean estimator of each  mid-energy $D_{j,k}$ groups from decomposition level $j.$   The GTME method is described in the following theorem:

\begin{thm}\label{thm g1}
Let $\hat{\mu}_{j,i}$ be the general trimean estimator based on $$\left\{D_{j,i},D_{j,i+M},D_{j,i+2M},...,D_{j,\left(N/2-M+i\right)}\right\}\coloneqq D(i,j)$$, where $D(i,j)$ for $1\le i\le M$ and $1\le j \le J$ is the $i$th group of  mid-energies  at level $j$ in a $J$-level NDWT of a fBm of size $N$ with Hurst exponent $H$. Then, the asymptotic distribution of  $\hat{\mu}_{j,i}$ is normal,
\begin{equation}
\hat{\mu}_{j,i}\widesim{\footnotesize{\mbox{approx}}} \mathcal{N}\left(c\left(\alpha,p\right)\lambda_j, \frac{2M}{N}f\left(\alpha,p\right)\lambda_j^{2}\right),
\end{equation}
where $$c\left(\alpha, p\right)= \frac{\alpha}{2}\log\left(\frac{1}{p\left(1-p\right)}\right)+\left(1-\alpha\right)\log2,$$ $$f\left(\alpha,p\right)= \frac{\alpha(1-2p)(\alpha-4p)}{4p(1-p)}+1, \mbox{and}$$
$$\lambda_j=\sigma^2\cdot2^{-\left(2H+1\right)j}.$$  In the previous, $\sigma$ is the standard deviation of wavelet coefficients from level 0. The Hurst exponent can be estimated as
\begin{equation}
\hat{H}=\frac{1}{M}\sum_{i=1}^M\hat{H}_i=\frac{1}{M}\sum_{i=1}^M\left(-\frac{\hat{\beta}_i}{2}-\frac{1}{2}\right)=-\frac{\bar{\beta}}{2}-\frac{1}{2},
\end{equation}
where $\bar{\beta}=\frac{1}{M}\sum_{i=1}^M\hat{\beta}_i$ is the average of regression slopes in the least square linear regression on pairs $\left(j, \log_2\left(\hat{\mu}_{j,i}\right)\right)$ over the $M$ groups, $i=1,2,...M$.
\end{thm}

The proof of Theorem $\ref{thm g1}$ is deferred to the appendix.
 
 To find the optimal $\alpha$ and $p$ minimizing the asymptotic variance of $\hat{\mu}_{j,i}$, we take partial derivatives of $f\left(\alpha, p\right)$ with respect to $\alpha$ and $p$ and set them to 0. The optimal $\hat{\alpha}$ and $\hat{p}$ can be obtained by solving
\begin{equation}
\begin{aligned}
&\frac{\partial f\left(\alpha,p\right)}{\partial\alpha}=-\frac{2p-1}{2p\left(1-p\right)}\alpha+\frac{1+p}{2\left(1-p\right)}-\frac{3}{2}=0, \\
&\frac{\partial f\left(\alpha,p\right)}{\partial p}=\frac{\alpha\left(2-\alpha\right)}{2\left(1-p\right)^2}+\frac{\alpha^2\left(2p-1\right)}{4p^2\left(1-p\right)^2}=0.\\
\end{aligned}
\end{equation}
Since $p\in\left(0,1/2\right)$, and $\alpha\in[0,1]$, we get the unique solution $p=1-\sqrt{2}/2\approx0.3$ and $\alpha=2p\approx0.6$. The Hessian matrix of $f\left(\alpha, p\right)$ is
\[\begin{bmatrix}
\frac{\partial^2f\left(\alpha,p\right)}{\partial\alpha^2} & \frac{\partial^2f\left(\alpha,p\right)}{\partial\alpha\partial p} \\
\frac{\partial^2f\left(\alpha,p\right)}{\partial\alpha\partial p} & \frac{\partial^2f\left(\alpha,p\right)}{\partial p^2} \\
 \end{bmatrix}=\]
\[ \begin{bmatrix}
-\frac{2p-1}{2p\left(1-p\right)} & \frac{2p^2-2\alpha p^2+\alpha\left(2p-1\right)}{2p^2\left(1-p\right)^2} \\
 \frac{2p^2-2\alpha p^2+\alpha\left(2p-1\right)}{2p^2\left(1-p\right)^2} & \frac{2p^3\alpha\left(2-\alpha\right)+\alpha^2 p\left(1-p\right)+\alpha^2\left(2p-1\right)^2}{2p^3\left(1-p\right)^3}\\
 \end{bmatrix}.\]
Since $-\frac{2p-1}{2p\left(1-p\right)}>0$ and the determinant is $5.66>0$ when $p=1-\sqrt{2}/2\approx0.3$ and $\alpha=2p\approx0.6$, the above Hessian matrix is positive definite. Therefore, $\hat{p}=1-\sqrt{2}/2$ and $\hat{\alpha}=2-\sqrt{2}$ provide the global minima of $f\left(\alpha, p\right)$, minimizing 
also the asymptotic variance of $\hat{\mu}_{j,i}$. In comparing these optimal $\hat{\alpha}\approx0.6$ and $\hat{p}\approx0.3$ with $\alpha=0.6$ and $p=1/3$ from the Gastwirth estimator, curiously, we find that the optimal general trimean estimator is very close to the Gastwirth estimator.

\subsection{General Trimean of the Logarithm of  Mid-energy  (GTLME) Method}
Previously discussed the GTME method calculates the general trimean estimator of the mid-energy first and then takes the logarithm. In this section, we will calculate the general trimean estimator of the logged  mid-energies  at each subgroup from level $j$. The following theorem describes the general trimean of the logarithm of  mid-energy,  the GTLME method.

\begin{thm}\label{thm g2}
Let $\hat{\mu}_{j,i}$ be the general trimean estimator based on $$\left\{\log\left(D_{j,i}\right),\log\left(D_{j,i+M}\right),...,\log\left(D_{j,\left(N/2-M+i\right)}\right)\right\}\coloneqq L(i,j)$$, where $L(i,j)$ is the $i$th group of logged  mid-energies'' at level $j$ in a $J$-level NDWT of a fBm of size $N$ with Hurst exponent $H$, $1\le i\le M$ and $1\le j \le J$. Then, the asymptotic distribution of  $\hat{\mu}_{j,i}$ is normal,
\begin{equation}
\hat{\mu}_{j,i}\widesim{\footnotesize{\mbox{approx}}} \mathcal{N}\left(c\left(\alpha,p\right)+\log\left(\lambda_j\right), \frac{2M}{N}f\left(\alpha, p\right)\right),
\end{equation}
where $$c\left(\alpha, p\right)=\frac{\alpha}{2}\log\left(\log\frac{1}{1-p}\cdot\log\frac{1}{p}\right)+\left(1-\alpha\right)\log\left(\log2\right),$$ and $$f\left(\alpha, p\right)=\frac{\alpha^2}{4g_1\left(p\right)}+\frac{\alpha\left(1-\alpha\right)}{2g_2\left(p\right)}+\frac{\left(1-\alpha\right)^2}{\left(\log2\right)^2},$$
$g_1\left(p\right)$ and $g_2\left(p\right)$ are two functions of $p$, $$\lambda_j=\sigma^2\cdot2^{-\left(2H+1\right)j},$$ and $\sigma^2$ is the variance of wavelet coefficients from level 0. The Hurst exponent can be estimated as
\begin{equation}
\hat{H}=\frac{1}{M}\sum_{i=1}^M\hat{H}_i=\frac{1}{M}\sum_{i=1}^M\left(\left(-\frac{\hat{\beta}_i}{2\log2}-\frac{1}{2}\right)\right)=-\frac{1}{2\log2}\bar{\beta}-\frac{1}{2},
\end{equation}
where $\bar{\beta}=\frac{1}{M}\sum_{i=1}^M\hat{\beta}_i$ is the average of regression slopes in the least square linear regressions on pairs $\left(j, \hat{\mu}_{j,i}\right)$ over the $M$ groups, $i=1,2,...M$.
\end{thm}

\begin{figure}[htb]
\centering
\includegraphics[width=4cm,height=4cm]{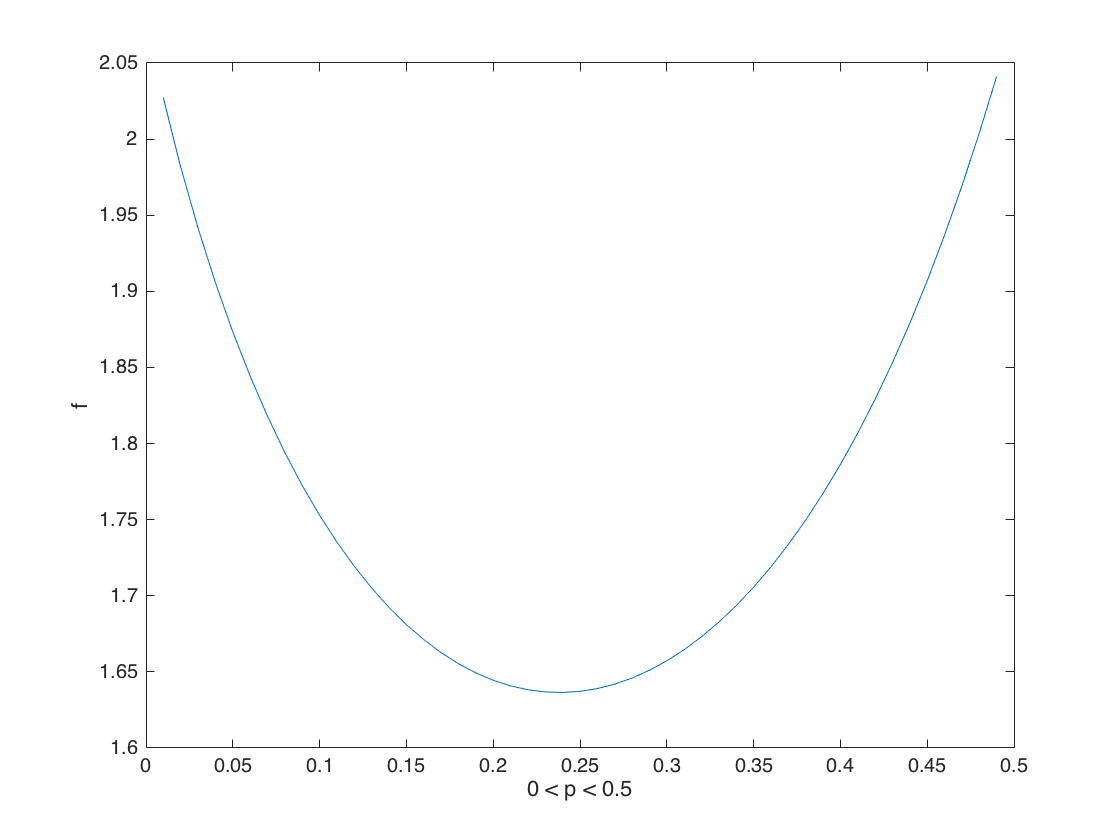}
\includegraphics[width=4cm,height=4cm]{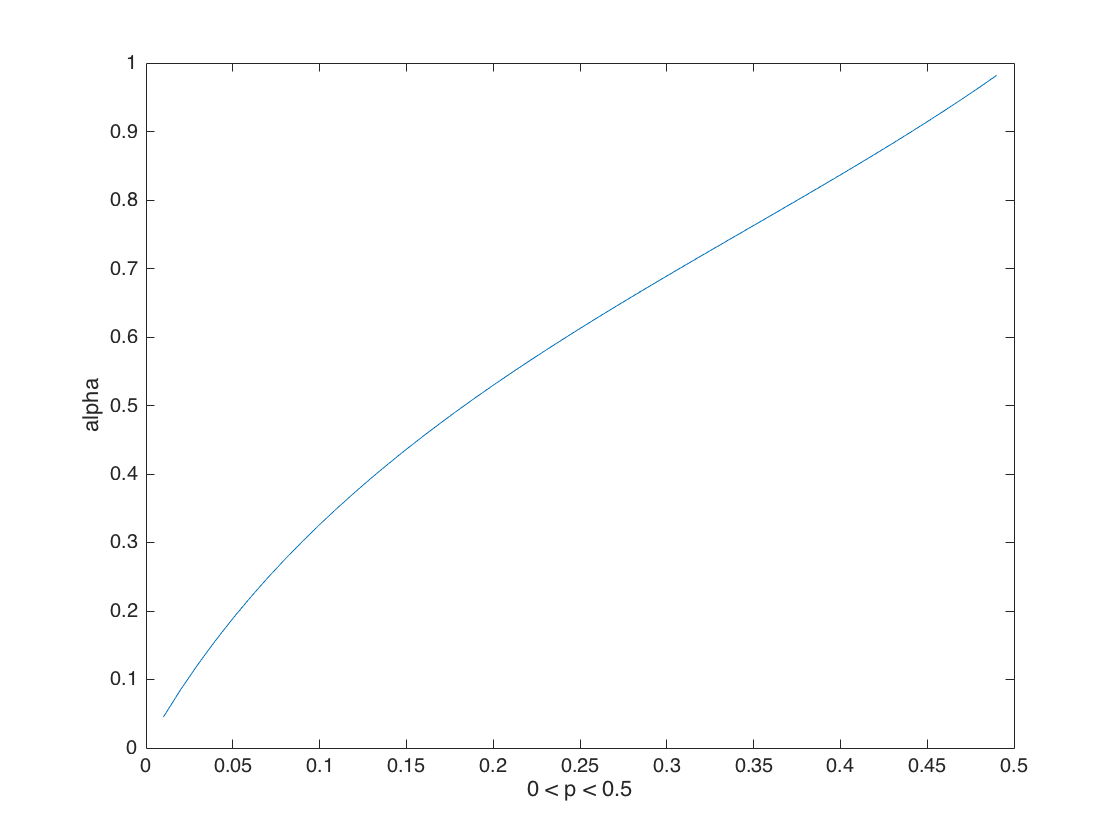}
\caption{Plot of $f\left(\alpha, p\right)$ against $p$ on the left; Plot of $\alpha$ against $p$ on the right}\label{Figure:1}
\end{figure}

The proof of Theorem $\ref{thm g2}$ is provided in the appendix. Similarly, as for the GTME, the optimal $\alpha$ and $p$ which minimize the asymptotic variance of $\hat{\mu}_{j,i}$ can be obtained by solving
\begin{equation}\label{eq: g2}
\frac{\partial f\left(\alpha,p\right)}{\partial\alpha}=0, \ \mbox{and}\ \frac{\partial f\left(\alpha,p\right)}{\partial p}=0.
\end{equation}
From the first equation in (\ref{eq: g2}) it can be derived that
\[\alpha=\frac{\frac{2}{\log\left( 2\right)^2}-\frac{1}{2}g_2\left(p\right)}{\frac{1}{2}g_1\left(p\right)-g_2\left(p\right)+\frac{2}{\left(\log2\right)^2}}.\]
The second equation in ($\ref{eq: g2}$) cannot be simplified to a finite form. As an illustration, we plot the $f\left(\alpha,p\right)$ with $p$ ranging from 0 to 0.5 and $\alpha$ being a function of $p$. The plot of $\alpha$ against $p$ is also shown in Figure $\ref{Figure:1}$. Numerical calculation gives $\hat{p}=0.24$ and $\hat{\alpha}=0.5965$. These optimal parameters are close to $\alpha=0.5$ and $p=0.25$ in the Tukey's trimean estimator, but  put  some more weight  on the median.

\subsection{Special Cases: Tukey's Trimean and Gastwirth Estimators}


The Tukey's trimean of the  mid-energy (TTME) method and Gastwirth of the mid-energy  (GME) method are described in the following Lemma.

\begin{Lemma}\label{thm1}
Let $\hat{\mu}_{j,i}^T$ and $\hat{\mu}_{j,i}^G$ be the Tukey's trimean and Gastwirth estimators based on
$D(i,j)$ defined in the Theorem $\ref{thm g1}$. Then the asymptotic distributions of  $\hat{\mu}_{j,i}^T$ and $\hat{\mu}_{j,i}^G$ are normal:
\begin{equation}
\hat{\mu}_{j,i}^T\widesim{\footnotesize{\mbox{approx}}} \mathcal{N}\left(c_1\lambda_j, \frac{5M}{3N}\lambda_j^2\right),
\end{equation}
\begin{equation}
\hat{\mu}_{j,i}^G\widesim{\footnotesize{\mbox{approx}}} \mathcal{N}\left(c_2\lambda_j, \frac{1.67M}{N}\lambda_j^2\right),
\end{equation}
where $c_1$ and $c_2$ are constants independent of $j$, $\lambda_j=\sigma^2\cdot2^{-\left(2H+1\right)j}$, and $\sigma^2$ is the variance of wavelet coefficients from level 0. The Hurst exponent can be estimated as
\begin{equation}
\hat{H}^T=-\frac{\bar{\beta}^T}{2}-\frac{1}{2}, \ \mbox{and}\ \hat{H}^G=-\frac{\bar{\beta}^G}{2}-\frac{1}{2},
\end{equation}
where $\bar{\beta}^T=\frac{1}{M}\sum_{i=1}^M\hat{\beta}^T_i$ and $\bar{\beta}^G=\frac{1}{M}\sum_{i=1}^M\hat{\beta}^G_i$ are the averages of regression slopes in the least square linear regression on pairs $\left(j, \log_2\left(\hat{\mu}_{j,i}^T\right)\right)$ and pairs $\left(j, \log_2\left(\hat{\mu}_{j,i}^G\right)\right)$ over the $M$ groups, respectively ($i=1,2,...M$).
\end{Lemma}

The following Lemma describes the Tukey's trimean (TTLME) and Gastwirth (GLME) of the logarithm of  mid-energy method.

\begin{Lemma}\label{thm3}
Let $\hat{\mu}_{j,i}^T$ and $\hat{\mu}_{j,i}^G$ be the Tukey's trimean estimator and Gastwirth estimator based on $L(i,j)$ defined in the Theorem $\ref{thm g2}$. The asymptotic distributions of  $\hat{\mu}_{j,i}^T$ and $\hat{\mu}_{j,i}^G$ are normal,
\begin{equation}
\hat{\mu}_{j,i}^T\widesim{\footnotesize{\mbox{approx}}} \mathcal{N}\left(-\left(2H+1\right)\log2j+c_3, V_1\right),
\end{equation}
\begin{equation}
\hat{\mu}_{j,i}^G\widesim{\footnotesize{\mbox{approx}}} \mathcal{N}\left(-\left(2H+1\right)\log2j+c_4, V_2\right),
\end{equation}
where $c_3$ ,$V_1$, $c_4$ and $V_2$ are constants independent of level $j$. The Hurst exponent can be estimated as
\begin{equation}
\hat{H^T}=-\frac{1}{2\log2}\bar{\beta}^T-\frac{1}{2},\ \mbox{and}\ \hat{H^G}=-\frac{1}{2\log2}\bar{\beta}^G-\frac{1}{2},
\end{equation}
where $\bar{\beta}^T=\frac{1}{M}\sum_{i=1}^M\hat{\beta}^T_i$ and $\bar{\beta}^G=\frac{1}{M}\sum_{i=1}^M\hat{\beta}^G_i$ are the averages of regression slopes in the least square linear regression on pairs $\left(j, \hat{\mu}_{j,i}^t\right)$ and pairs $\left(j, \hat{\mu}_{j,i}^g\right)$ ,$i=1,2,...M$, over the $M$ groups.
\end{Lemma}

The proofs of Lemma $\ref{thm1}$ and Lemma $\ref{thm3}$ are provided in the appendix.

\section{Simulation}
\begin{figure}[htb]
\centering
\includegraphics[width=9cm,height=4cm]{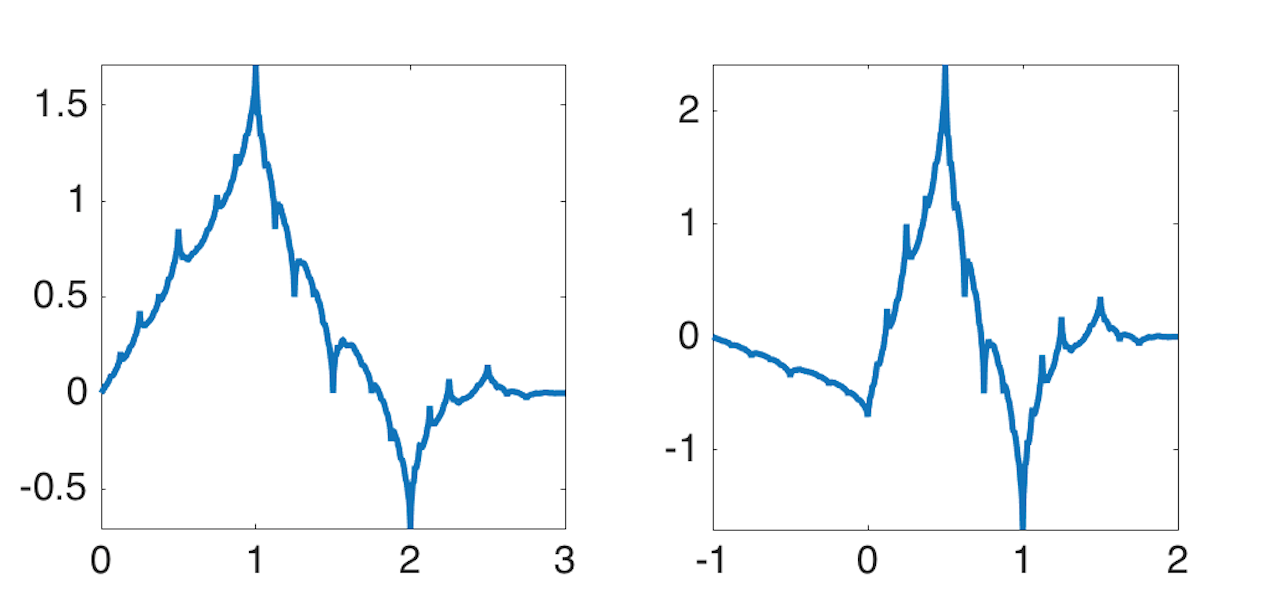}
\caption{Pollen scaling and wavelet functions for $\pi/4$}\label{Figure:2}
\end{figure}

We simulate fractional Brownian motion (fBm) signals of sizes $N=2^{10}, N=2^{11}$, and $N=2^{12}$ with Hurst exponent $H=0.3, 0.5, 0.7, 0.8, 0.9$, respectively. NDWT of depth $J=10$ using Pollen wavelets with angles $\pi/6$ (Daubechies 2), $\pi/4$, $\pi/3$, and $\pi/2$ (Haar) are performed on each simulated signal to obtain wavelet coefficients.  Pollen generates a family possessing continuum many wavelet bases of various degrees of regularity \cite{vidakovic2002pollen}. Special cases of Pollen's representation for $\pi/6$ and $\pi/2$ give Daubechies 2 filter and Haar filter, respectively. Figure $\ref{Figure:2}$ depicts scaling and wavelet functions for $\pi/4$.

The proposed methods (with 6 variations) are applied on the NDWT coefficients to estimate Hurst exponent $H$. Coefficients on each level are divided into eight groups ($M=8$) for all proposed methods, and we use wavelet coefficients from levels 4 to 10 for the least square linear regression. The estimation performance of the proposed methods are compared to four other existing methods: Veitch $\&$ Abry (VA) method, Soltani, Simard, $\&$ Boichu (SSB) method, MEDL method, and MEDLA method. The GTME and GTLME methods are based on the optimal parameters to minimize the variances. Estimation performances are compared in terms of mean, variance, and mean square error (MSE) based on 300 repetitions for each case.

The proposed methods preform the best using Haar wavelet (Pollen wavelets with angle $\pi/2$), and the simulation results are shown in Table $\ref{table:1}$ to Table $\ref{table:3}$ for fBm of sizes $N=2^{10}, N=2^{11}$, and $N=2^{12}$, respectively. Similar results are obtained for other wavelets. For each $H$ (corresponding to each row in the table), the smallest variances and MSEs are highlighted in bold. From simulations results, at least one of our 6 variations outperforms MEDL and MEDLA for all $H$ and fBm of all three sizes. Compared with VA and SSB methods, our methods yield significantly smaller variances and MSEs when $H>0.5$ for fBm of all three sizes. When $H=0.3\ \mbox{and}\ 0.5$, our methods are still comparable to VA and SSB. Although the performances of our 6 variations are very similar regarding to variances and MSEs, the TTME method based on Tukey's trimean estimator of the  mid-energy  has the best performance among all of them. As expected, the variances of GTME based on the optimal parameters are smaller than or equal to those of GME and TTME methods in most cases, and similarly, in most cases the optimized GTLME method is superior to other logged  mid-energy  methods TTLME and GLME with respect to variances. However, this superiority is not significant, since the variances of all six proposed methods are close to each other.

\begin{table*}[h]
\centering
\caption{Simulation Results for $N=2^{10}$ fBm using Haar wavelet (300 Replications)}\label{table:1}
\begin{tabular*}{1\textwidth}{@{\extracolsep{\fill}}ccccccccccc@{}}
\hline
\multicolumn{6}{r}{Methods} \\ \cmidrule(r){2-11}
$H$ & VA & SSB & MEDL & MEDLA & TTME & GME & TTLME & GLME & GTME & GTLME\\
\hline
\multicolumn{11}{c}{\multirow{2}*{$\hat{H}$}}\\
\\
\hline
0.3 & 0.2429 & 0.2429 & 0.2368 & 0.2365 & 0.2397 & 0.2393 & 0.2393 & 0.2390 & 0.2395 & 0.2396 \\
0.5 & 0.4482 & 0.4944 & 0.4800 & 0.4821 & 0.4893 & 0.4870 & 0.4872 & 0.4861 & 0.4882 & 0.4876 \\
0.7 & 0.5209 & 0.7232 & 0.6996 & 0.7002 & 0.7094 & 0.7072 & 0.7067 & 0.7058 & 0.7085 & 0.7071 \\
0.8 & 0.5171 & 0.8304 & 0.7999 & 0.7991 & 0.8050 & 0.8020 & 0.8031 & 0.8008 & 0.8037 & 0.8044 \\
0.9 & 0.4976 & 0.9410 & 0.9020 & 0.8951 & 0.9016 & 0.9004 & 0.9024 & 0.9009 & 0.9012 & 0.9032  \\
\hline
\multicolumn{11}{c}{\multirow{2}*{Variances}}\\
\\
\hline
0.3 & 0.0024 & 0.0024 & 0.0025 & 0.0022 & $\textbf{0.0020}$ & 0.0021 & 0.0022 & 0.0022 & 0.0021 & 0.0022 \\
0.5 & $\textbf{0.0018}$ & 0.0034 & 0.0035 & 0.0030 & 0.0030 & 0.0031 & 0.0031 & 0.0031 & 0.0030 & 0.0030 \\
0.7 & 0.0048 & 0.0055 & 0.0049 & 0.0038 & $\textbf{0.0036}$ & $\textbf{0.0036}$ & 0.0038 & 0.0037 & $\textbf{0.0036}$& 0.0038 \\
0.8 & 0.0067 & 0.0071 & 0.0061 & 0.0044 & $\textbf{0.0041}$ & 0.0044 & 0.0047 & 0.0047 & 0.0042 & 0.0047 \\
0.9 & 0.0064 & 0.0085 & 0.0066 & 0.0047 & $\textbf{0.0041}$ & 0.0043 & 0.0047 & 0.0046 & 0.0042 & 0.0047 \\
\hline
\multicolumn{11}{c}{\multirow{2}*{MSEs}}\\
\\
\hline
0.3 & 0.0057 & $\textbf{0.0056}$ & 0.0065 & 0.0062 & $\textbf{0.0056}$ & 0.0058 & 0.0059 & 0.0059 & 0.0057 & 0.0058 \\
0.5 & 0.0044 & 0.0034 & 0.0038 & 0.0034 & $\textbf{0.0031}$ & 0.0032 & 0.0032 & 0.0033 & $\textbf{0.0031}$ & 0.0032 \\
0.7 & 0.0368 & 0.0060 & 0.0049 & 0.0038 & 0.0037 & $\textbf{0.0036}$ & 0.0038 & 0.0037 & 0.0037 & 0.0039 \\
0.8 & 0.0867 & 0.0080 & 0.0061 & 0.0044 & $\textbf{0.0042}$ & 0.0044 & 0.0047 & 0.0047 & $\textbf{0.0042}$ & 0.0047 \\
0.9 & 0.1683 & 0.0101 & 0.0066 & 0.0047 & $\textbf{0.0041}$ & 0.0043 & 0.0047 & 0.0046 & 0.0042 & 0.0047 \\
\hline
\end{tabular*}
\end{table*}

\begin{table*}[h]
\centering
\caption{Simulation Results for $N=2^{11}$ fBm using Haar wavelet (300 Replications)}\label{table:2}
\begin{tabular*}{1\textwidth}{@{\extracolsep{\fill}}ccccccccccc@{}}
\hline
\multicolumn{6}{r}{Methods} \\ \cmidrule(r){2-11}
$H$ & VA & SSB & MEDL & MEDLA & TTME & GME & TTLME & GLME & GTME & GTLME\\
\hline
\multicolumn{11}{c}{\multirow{2}*{$\hat{H}$}}\\
\\
\hline
0.3 & 0.2448 & 0.2420 & 0.2385 & 0.2382 & 0.2393 & 0.2389 & 0.2392 & 0.2390 & 0.2392 & 0.2394 \\
0.5 & 0.4548 & 0.4822 & 0.4721 & 0.4725 & 0.4755 & 0.4745 & 0.4743 & 0.4740 & 0.4749 & 0.4744 \\
0.7 & 0.5216 & 0.7125 & 0.6917 & 0.6931 & 0.6977 & 0.6963 & 0.6959 & 0.6954 & 0.6969 & 0.6963 \\
0.8 & 0.5224 & 0.8201 & 0.7937 & 0.7933 & 0.7986 & 0.7972 & 0.7979 & 0.7970 & 0.7979 & 0.7985 \\
0.9 & 0.5028 & 0.9301 & 0.8988 & 0.8922 & 0.8958 & 0.8957 & 0.8976 & 0.8965 & 0.8958 & 0.8982 \\
\hline
\multicolumn{11}{c}{\multirow{2}*{Variances}}\\
\\
\hline
0.3 & 0.0013 & 0.0011 & 0.0011 & $\textbf{0.0010}$ & $\textbf{0.0010}$ & $\textbf{0.0010}$ & $\textbf{0.0010}$ & $\textbf{0.0010}$ & $\textbf{0.0010}$ & $\textbf{0.0010}$ \\
0.5 & $\textbf{0.0006}$ & 0.0013 & 0.0015 & 0.0011 & 0.0010 & 0.0011 & 0.0011 & 0.0012 & 0.0011 & 0.0011 \\
0.7 & 0.0041 & 0.0022 & 0.0022 & 0.0016 & $\textbf{0.0015}$ & $\textbf{0.0015}$ & 0.0016 & 0.0016 & $\textbf{0.0015}$ & 0.0016 \\
0.8 & 0.0071 & 0.0039 & 0.0035 & 0.0026 & $\textbf{0.0023}$ & 0.0025 & 0.0026 & 0.0026 & 0.0024 & 0.0026 \\
0.9 & 0.0065 & 0.0053 & 0.0046 & 0.0032 & $\textbf{0.0027}$ & 0.0029 & 0.0032 & 0.0032 & 0.0028 & 0.0032 \\
\hline
\multicolumn{11}{c}{\multirow{2}*{MSEs}}\\
\\
\hline
0.3 & $\textbf{0.0044}$ & $\textbf{0.0044}$ & 0.0049 & 0.0048 & 0.0046 & 0.0047 & 0.0047 & 0.0047 & 0.0046 & 0.0046 \\
0.5 & 0.0027 & $\textbf{0.0016}$ & 0.0023 & 0.0019 & $\textbf{0.0016}$ & 0.0018 & 0.0018 & 0.0018 & 0.0017 & 0.0018 \\
0.7 & 0.0359 & 0.0024 & 0.0023 & 0.0016 & $\textbf{0.0015}$ & 0.0016 & 0.0016 & 0.0017 & $\textbf{0.0015}$ & 0.0016 \\
0.8 & 0.0841 & 0.0042 & 0.0035 & 0.0027 & $\textbf{0.0023}$ & 0.0025 & 0.0026 & 0.0026 & 0.0024 & 0.0026 \\
0.9 & 0.1642 & 0.0062 & 0.0046 & 0.0033 & $\textbf{0.0027}$ & 0.0029 & 0.0032 & 0.0032 & 0.0028 & 0.0032 \\
\hline
\end{tabular*}
\end{table*}

\begin{table*}[h]
\centering
\caption{Simulation Results for $N=2^{12}$ fBm using Haar wavelet (300 Replications)}\label{table:3}
\begin{tabular*}{1\textwidth}{@{\extracolsep{\fill}}ccccccccccc@{}}
\hline
\multicolumn{6}{r}{Methods} \\ \cmidrule(r){2-11}
$H$ & VA & SSB & MEDL & MEDLA & TTME & GME & TTLME & GLME & GTME & GTLME\\
\hline
\multicolumn{11}{c}{\multirow{2}*{$\hat{H}$}}\\
\\
\hline
0.3 & 0.2427 & 0.2364 & 0.2337 & 0.2339 & 0.2342 & 0.2339 & 0.2341 & 0.2339 & 0.2341 & 0.2343 \\
0.5 & 0.4585 & 0.4759 & 0.4672 & 0.4679 & 0.4701 & 0.4691 & 0.4693 & 0.4686 & 0.4695 & 0.4695 \\
0.7 & 0.5228 & 0.7002 & 0.6843 & 0.6857 & 0.6890 & 0.6882 & 0.6879 & 0.6877 & 0.6885 & 0.6879 \\
0.8 & 0.5124 & 0.8104 & 0.7925 & 0.7907 & 0.7932 & 0.7928 & 0.7930 & 0.7928 & 0.7929 & 0.7932 \\
0.9 & 0.4935 & 0.9190 & 0.8976 & 0.8928 & 0.8932 & 0.8935 & 0.8951 & 0.8945 & 0.8932 & 0.8954  \\
\hline
\multicolumn{11}{c}{\multirow{2}*{Variances}}\\
\\
\hline
0.3 & 0.0009 & $\textbf{0.0006}$ & 0.0007 & $\textbf{0.0006}$ & $\textbf{0.0006}$ & $\textbf{0.0006}$ & $\textbf{0.0006}$ & $\textbf{0.0006}$ & $\textbf{0.0006}$ & $\textbf{0.0006}$ \\
0.5 & $\textbf{0.0004}$ & 0.0007 & 0.0008 & 0.0006 & 0.0005 & 0.0006 & 0.0006 & 0.0006 & 0.0006 & 0.0006 \\
0.7 & 0.0046 & 0.0013 & 0.0012 & 0.0010 & $\textbf{0.0009}$ & 0.0010 & 0.0010 & 0.0010 & 0.0010 & 0.0010 \\
0.8 & 0.0060 & 0.0019 & 0.0018 & 0.0014 & $\textbf{0.0013}$ & 0.0014 & 0.0014 & 0.0015 & $\textbf{0.0013}$ & 0.0014 \\
0.9 & 0.0049 & 0.0031 & 0.0028 & 0.0021 & $\textbf{0.0019}$ & 0.0020 & 0.0021 & 0.0021 & $\textbf{0.0019}$ & 0.0021 \\
\hline
\multicolumn{11}{c}{\multirow{2}*{MSEs}}\\
\\
\hline
0.3 & $\textbf{0.0042}$ & 0.0046 & 0.0051 & 0.0050 & 0.0049 & 0.0050 & 0.0049 & 0.0050 & 0.0049 & 0.0049 \\
0.5 & 0.0021 & $\textbf{0.0012}$ & 0.0019 & 0.0016 & 0.0014 & 0.0015 & 0.0016 & 0.0016 & 0.0015 & 0.0015 \\
0.7 & 0.0360 & 0.0013 & 0.0015 & 0.0012 & $\textbf{0.0011}$ & $\textbf{0.0011}$ & 0.0012 & 0.0012 & $\textbf{0.0011}$ & 0.0012 \\
0.8 & 0.0887 & 0.0020 & 0.0019 & 0.0015 & $\textbf{0.0013}$ & 0.0014 & 0.0015 & 0.0015 & 0.0014 & 0.0015 \\
0.9 & 0.1702 & 0.0035 & 0.0028 & 0.0021 & $\textbf{0.0019}$ & 0.0020 & 0.0021 & 0.0021 & $\textbf{0.0019}$ & 0.0021 \\
\hline
\end{tabular*}
\end{table*}

\section{Application}
In this section, we apply the proposed methodology to high frequency measurements in pupillary response behavior (PRB). PRB refers to changes in pupil diameter in response to simple or complex stimuli. Participants in this study included 24 older adults, solicited from the patient pool of the Bascom Palmer Eye Institute of the University of Miami School of Medicine. Participants were selected on the basis of having either no ocular disease or only Age-related Macular Degeneration (AMD), as assessed by patient history and clinical testing. Participants were selected based on the diagnosis of AMD and their best eye near distance (40 cm) visual acuity.

Participants were assigned to 3 groups ( one control group and 2 experimental groups). The control group is a set of individuals with healthy, unaffected vision and no evidence of any ocular disease or trauma. Individuals in 2 experimental groups had varying visual acuity and were diagnosed with AMD. The number of participants is 6 in control group, 12 in group I, and 6 in group II. The data analysis process consists of first cleaning the data by removing the blink and equipment artifacts and then segmenting the data stream for each participant into equal length pieces of 2048 observations. The number of 2048 length data sets that were extracted from the collective data sets of the individuals within each group are shown in Table $\ref{table:group}$.

Traditional statistical methods have not been successfully used for examining the PRB of older adults or individuals with visual impairments. Researchers have utilized simple statistical methods for analyzing PRB, for example, comparing the relative mean or variance of pupil size deviation in response to stimuli; some sophisticated techniques have also been utilized, like power, frequency and spectral analysis using mathematical tools. But most of them failed to characterize the underlying patterns within time series PRB data. Wavelet analysis to estimate the Hurst exponent of the high-frequency, time series physiological data is a useful tool for detecting these hidden patterns and differentiating individuals based on these unique patterns in their physiological behavior.

Table $\ref{table:pupilH}$ provides descriptive statistics of the estimated Hurst exponent $\hat{H}$ in each group using our proposed methods and four standard methods to compare with. As can be seen, there are clear monotonic trends for $\hat{H}$ across the participant groups. The control group exhibited the smallest value for $\hat{H}$, followed by group I and group II, and those monotonic trends are presented in both the mean and median values. In fact, signals with smaller Hurst exponent $H$ tend to be more disordered and unsystematic, therefore the monotonic trends indicating that individuals with higher severity of visual impairment have less disordered pupil diameter signals. Similar results for original data without being pre-processed are shown in Table $\ref{table:pupil 1}$. No monotonic trends of $\hat{H}$ across groups have been observed in original data, indicating the blinks indeed add noise to the data.

Like in many human-subject studies, the limited number of participants is a major disadvantage, but in PRB data set, each subject has enough measurements to segment into multiple pieces with a length of 2048 observations. This induces dependence between the runs, and hierarchical models accommodating for the subject induced dependence are needed. If we denote $i$ to be the group index where the piece of observations is from, with $i=0$ for control group, $i=1$ for group I, $i=2$ for group II, and $n_j$ as the number of pieces generated from subject $j$ (j=1,2,..,24), the estimated Hurst exponent $\hat{H}_{ijk}$ for the $k$th piece of subject $j$ nested in group $i$ can be expressed in the following model:
\begin{equation}
\hat{H}_{ijk}=\mu+\alpha_i+\beta_{j\left(i\right)}+\epsilon_{ijk},
\end{equation}
where $\mu$ is the overall mean, $\alpha_i$ is the effect for $i$th group, $\beta_{j\left(i\right)}$ is the effect for $j$th participant within $i$th group, and $\epsilon_{ijk}$ is the random error. The objective is to classify the groups based on the estimated Hurst exponent for a given pupil diameter data. In avoid of dependency between data sets due to subject effects, the estimated $\hat{\beta}_{j\left(i\right)}$ is first subtracted from $\hat{H}_{ijk}$, and multinomial logistic regression model is fitted on the data $\left\{\left(\hat{H}_{ijk}-\hat{\beta}_{j\left(i\right)}, i\right), i=0,1,2, j=1,2,...,24, k=1,2,...,n_j\right\}$. To test the model performance, we randomly choose $80\%$ of the data points to form a training set, and the remaining forms the testing set. Model is developed on the training set and applied on the testing set; misclassification rate is reported in Table $\ref{table:ce}$.

\begin{table*}[h]
\centering
\begin{threeparttable}
\caption{Group characterization summary}\label{table:group}
\begin{tabular*}{1\textwidth}{@{\extracolsep{\fill}}cccccc@{}}
\hline
Group & N & Visual Acuity & AMD &  Number of preprocessed data sets\\
\hline
Control & 6 & 20/20-20/40 & No & 49\\
I & 12 & 20/20-20/100 & Yes  & 184\\
II & 6 & 20/100 & Yes  & 170 \\
\hline
\end{tabular*}
 \begin{tablenotes}
      \footnotesize
      \item Note. N represents the number of individuals in the group; Visual Acuity represents the range of Snellen acuity scores for the individuals in the given group, AMD represents whether the individuals were diagnosed with age-related macular degeneration or not, and Number of cleaned data sets shows the number of 2048 length data sets extracted from individuals within each group.
    \end{tablenotes}
\end{threeparttable}
\end{table*}

\begin{table*}[h]
\centering
\caption{Descriptive Statistics Group Summary (Blinks removed)}\label{table:pupilH}
\begin{tabular*}{1\textwidth}{@{\extracolsep{\fill}}ccccccccccc@{}}
\hline
\multicolumn{6}{r}{Methods} \\ \cmidrule(r){2-11}
Group & Veitch and Abry & Soltani & MEDL & MEDLA & TTME & GME & TTLME & GLME & GTME & GTLME\\
\hline
\multicolumn{11}{c}{\multirow{2}*{Mean of $\hat{H}$}}\\
\\
\hline
Control & 0.1495 & 0.4103 & 0.2741 & 0.2501 & 0.2416 & 0.2505 & 0.2537 & 0.2559 & 0.2467 & 0.2530 \\
I & 0.1855 & 0.4210 & 0.2733 & 0.2727 & 0.2638 & 0.2686 & 0.2716 & 0.2738 & 0.2666 & 0.2705 \\
II & 0.1682 & 0.5464 & 0.3597 & 0.3202 & 0.2984 & 0.3116 & 0.3257 & 0.3231 & 0.3057 & 0.3286 \\
\hline
\multicolumn{11}{c}{\multirow{2}*{Median of $\hat{H}$}}\\
\\
\hline
Control & 0.1474 & 0.4088 & 0.2869 & 0.2596 & 0.2535 & 0.2609 & 0.2625 & 0.2683 & 0.2577 & 0.2615 \\
I & 0.2073 & 0.4331 & 0.3033 & 0.2951 & 0.2938 & 0.2990 & 0.3054 & 0.3038 & 0.2976 & 0.3042 \\
II & 0.1814 & 0.5436 & 0.3514 & 0.3240 & 0.2973 & 0.3127 & 0.3261 & 0.3313 & 0.3058 & 0.3254 \\
\hline
\multicolumn{11}{c}{\multirow{2}*{Variance of $\hat{H}$}}\\
\\
\hline
Control & 0.0084 & 0.0195 & 0.0088 & 0.0086 & 0.0091 & 0.0094 & 0.0092 & 0.0094 & 0.0095 & 0.0092 \\
I & 0.0120 & 0.0421 & 0.0213 & 0.0206 & 0.0191 & 0.0196 & 0.0200 & 0.0204 & 0.0194 & 0.0198 \\
II & 0.0114 & 0.0373 & 0.0428 & 0.0066 & 0.0061 & 0.0060 & 0.0087 & 0.0065 & 0.0060 & 0.0102 \\
\hline
\end{tabular*}
\end{table*}

\begin{table*}[h]
\centering
\footnotesize
\caption{Descriptive Statistics Group Summary (Blinks not removed)}\label{table:pupil 1}
\begin{tabular*}{1\textwidth}{@{\extracolsep{\fill}}ccccccccccc@{}}
\hline
\multicolumn{6}{r}{Methods} \\ \cmidrule(r){2-11}
Group & Veitch and Abry & Soltani & MEDL & MEDLA & TTME & GME & TTLME & GLME & GTME & GTLME\\
\hline
\multicolumn{11}{c}{\multirow{2}*{Mean of $\hat{H}$}}\\
\\
\hline
Control & 0.1892 & 0.4627 & 0.3582 & 0.3497 & 0.3394 & 0.3498 & 0.3473 & 0.3498 & 0.3459 & 0.3446 \\
I & 0.1812 & 0.4272 & 0.2860 & 0.2918 & 0.2858 & 0.2900 & 0.2900 & 0.2923 & 0.2891 & 0.2888 \\
II & 0.1565 & 0.5510 & 0.3672 & 0.3295 & 0.3113 & 0.3262 & 0.3373 & 0.3360 & 0.3201 & 0.3416 \\
\hline
\multicolumn{11}{c}{\multirow{2}*{Median of $\hat{H}$}}\\
\\
\hline
Control & 0.2144 & 0.5194 & 0.3502 & 0.3574 & 0.3464 & 0.3559 & 0.3513 & 0.3542 & 0.3510 & 0.3508 \\
I & 0.1867 & 0.4467 & 0.3072 & 0.3091 & 0.2943 & 0.2981 & 0.3100 & 0.3055 & 0.3014 & 0.3080 \\
II & 0.1758 & 0.5528 & 0.3568 & 0.3279 & 0.2999 & 0.3219 & 0.3330 & 0.3345 & 0.3104 & 0.3320 \\
\hline
\multicolumn{11}{c}{\multirow{2}*{Variance of $\hat{H}$}}\\
\\
\hline
Control & 0.0183 & 0.0311 & 0.0294 & 0.0288 & 0.0258 & 0.0270 & 0.0253 & 0.0261 & 0.0266 & 0.0246 \\
I & 0.0184 & 0.0449 & 0.0245 & 0.0245 & 0.0250 & 0.0251 & 0.0244 & 0.0248 & 0.0252 & 0.0242  \\
III & 0.0212 & 0.0363 & 0.0420 & 0.0082 & 0.0098 & 0.0089 & 0.0102 & 0.0083 & 0.0093 & 0.0132 \\
\hline
\end{tabular*}
\end{table*}

\begin{table*}[h]
\centering
\footnotesize
\caption{Classification error}\label{table:ce}
\begin{tabular*}{1\textwidth}{@{\extracolsep{\fill}}ccccccccccc@{}}
\hline
 &Veitch and Abry & Soltani & MEDL & MEDLA & TTME & GME & TTLME & GLME & GTME & GTLME\\
\hline
Original data &0.3837 & 0.3721 & 0.3605 & 0.3372 & 0.3721 & 0.3721 & 0.3721 & 0.3721 & 0.3721 & 0.3837 \\
Blinks removed & 0.2840 & 0.3210 & 0.2469 & 0.2963 & 0.2469 & 0.2346 & 0.2593 & 0.2346 & 0.2469 & 0.2716  \\
\hline
\end{tabular*}
\end{table*}

\section{Conclusions}
In this paper, we proposed methodologies and derived 6 variations to improve the robustness of estimation of Hurst exponent $H$ in one-dimensional setting. Non-decimated wavelet transforms (NDWT) are utilized for its redundancy and time-invariance. Instead of using mean or median of the derived distribution on level-wise wavelet coefficients, we defined the general trimean estimators that combine the median's emphasis on center values with the quantiles' attention to the extremes and used them on the level-wise derived distributions to estimate $H$.

The proposed variations were: 1) general trimean of the  mid-energy  (GTME) method; 2) general trimean of the logarithm of  mid-energy  (GTLME) method; 3) Tukey's trimean of the  mid-energy (TTME) method; 4) Tukey's trimean of the logged  mid-energy  (TTLME) method; 5) Gastwirth of the  mid-energy  (GME) method; 6) Gastwirth of the logged  mid-energy  (GLME) method. The GTME and GTLME methods are based on the derived optimal parameters in general trimean estimators to minimize the estimation variances. Tukey's trimean and Gastwirth estimators are two special cases following the general trimean estimators' framework. These estimators are applied on both  mid-energy  
(as defined by Soltani et al.,\cite{soltani2004estimation}) and logarithm of the  mid-energy at each NDWT level. The estimation performance of the proposed methods are compared to four other existing methods: Veitch $\&$ Abry (VA) method, Soltani, Simard, $\&$ Boichu (SSB) method, MEDL method, and MEDLA method.

Simulation results indicate our proposed methods preform the best using Haar wavelet. We found that at least one of our 6 variations outperforms MEDL and MEDLA for all $H$ and fBm of all three sizes. Compared with VA and SSB methods, our methods yield significantly smaller variances and MSEs when $H>0.5$ for fBm of all three sizes. When $H=0.3\ \mbox{and}\ 0.5$, our methods are still comparable to VA and SSB. Although the performances of our 6 variations are very similar regarding to variances and MSEs, the TTME method based on Tukey's trimean estimator of the  mid-energy has the best performance among all of them.

The proposed methods have been applied to a real pupillary response behavior (PRB) data set to extract meaningful characteristics from the PRB data of older adults with and without visual impairment. Estimated Hurst exponents base on wavelet analysis capture the unique pattern of a data signal that cannot be represented by the the trends and traditional statistical summaries of the signal in that the magnitude of the pupil diameter depends on ambient light, not on the inherent eye function. Our proposed methods helped to detect those unique patterns and differentiate individuals based on the estimated Hurst parameters $\hat{H}$. Monotonic trends have been found for $\hat{H}$ across the participant groups, and individuals with higher severity of visual impairment have more regular pupil diameter signals. This increase of regularity with increase of the degree of pathology is common for many other biometric signals: EEG, EKG, high frequency protein mass-spectra, high resolution medical images of tissue, to list a few.

\appendices
\section{Appendix: Technical Proofs.}
\noindent
{\bf Proof of Theorem~\ref{thm g1}.}
\begin{proof}
A single wavelet coefficient in a non-decimated wavelet transform of a fBm of size $N$ with Hurst exponent $H$ is normally distributed, with variance depending on its level $j$, therefore, each pair $d_{j,k}$ and $d_{j,k+N/2}$ in mid-energy $D_{j,k}$ are assumed to be independent and follow the same normal distribution.
\[d_{j,k},\ d_{j,k+N/2}\sim \mathcal{N}\left(0, 2^{-\left(2H+1\right)j}\sigma^2\right).\]
Then the  mid-energy  is defined as
\[D_{j,k}=\frac{\left(d_{j,k}^2+d_{j,k+N/2}^2\right)}{2},\ j=1,..,J,\mbox{and}\ k=1,...,N/2,\]
and it can be readily shown that $D_{j,k}$ has exponential distribution
 with scale parameter $\lambda_j=\sigma^2\cdot2^{-\left(2H+1\right)j}$, i.e.,
\begin{equation*}
f\left(D_{j,k}\right)=\lambda_j^{-1}e^{-\lambda_j^{-1}D_{j,k}},\ \mbox{for any}\ k=1,..,N/2.
\end{equation*}
Therefore the $i$th subgroup $\left\{D_{j,i},D_{j,i+M},D_{j,i+2M},...,D_{j,\left(N/2-M+i\right)}\right\}$ are i.i.d. $\exp\left(\lambda_j^{-1}\right)$, and when applying general trimean estimator $\hat{\mu}_{j,i}$ on $\left\{D_{j,i},D_{j,i+M},D_{j,i+2M},...,D_{j,\left(N/2-M+i\right)}\right\}$, following the derivation in Section $\ref{section general}$, we have
\[\bm{\xi}=[\log\left(\frac{1}{1-p}\right)\lambda_j\ \ \ \log\left(2\right)\lambda_j\ \ \ \log\left(\frac{1}{p}\right)\lambda_j]^T,\]
and
\[\Sigma=\begin{bmatrix}
\frac{p}{\left(1-p\right)\lambda_j^2} & \frac{p}{\left(1-p\right)\lambda_j^2} & \frac{p}{\left(1-p\right)\lambda_j^2} \\
\frac{p}{\left(1-p\right)\lambda_j^2} & \frac{1}{\lambda_j^2} & \frac{1}{\lambda_j^2} \\
\frac{p}{\left(1-p\right)\lambda_j^2} & \frac{1}{\lambda_j^2} & \frac{1-p}{p\lambda_j^2} \\
 \end{bmatrix}_{3\times 3},\]
therefore, the asymptotic distribution of  $\hat{\mu}_{j,i}$ is normal with mean
\begin{equation*}
\begin{aligned}
\mathbb{E}\left(\hat{\mu}_{j,i}\right)&=A\cdot\bm{x}\\
&=\left(\frac{\alpha}{2}\log\left(\frac{1}{p\left(1-p\right)}\right)+\left(1-\alpha\right)\log2\right)\lambda_j \\
&\triangleq c\left(\alpha, p\right)\lambda_j,\\
\end{aligned}
\end{equation*}
and variance
\begin{equation*}
\begin{aligned}
\Var\left(\hat{\mu}_{j,i}\right)&=\frac{2M}{N}A\Sigma A^T \\
&=\frac{2M}{N}\left(\frac{\alpha(1-2p)(\alpha-4p)}{4p(1-p)}+1\right)\lambda_j^{2}\\
&\triangleq\frac{2M}{N}f\left(\alpha,p\right)\lambda_j^{2}.\\
\end{aligned}
\end{equation*}
\end{proof}

\noindent
{\bf Proof of Theorem~$\ref{thm g2}$.}
\begin{proof}
We have stated that $D_{j,k}\sim {\cal E}xp\left(\lambda_j^{-1}\right)$ with scale parameter $\lambda_j=\sigma^2\cdot2^{-\left(2H+1\right)j}$, so that
\begin{equation*}
f\left(D_{j,k}\right)=\lambda_j^{-1}e^{-\lambda_j^{-1}D_{j,k}},\ \mbox{for any}\ k=1,..,N/2.
\end{equation*}
Let $y_{j,k}=\log\left(D_{j,k}\right)$ for any $j=1,...,J$ and $k=1,...,N/2$. The pdf and cdf of $y_{j,k}$ are
\[f\left(y_{j,k}\right)=\lambda_j^{-1}e^{-\lambda_j^{-1}y_{j,k}}e^{y_{j,k}},\]
and
\[F\left(y_{j,k}\right)=1-e^{-\lambda_j^{-1}e^{y_{j,k}}}.\]
The $p$-quantile can be obtained by solving $F\left(y_p\right)=1-e^{-\lambda_j^{-1}e^{y_{j,k}}}=p$, and $y_p=\log\left(-\lambda_j\log\left(1-p\right)\right)$. Then it can be shown that $f\left(y_p\right)=-\left(1-p\right)\log\left(1-p\right)$. When applying the general trimean estimator $\hat{\mu}_{j,i}$ on $\left\{\log\left(D_{j,i}\right),\log\left(D_{j,i+M}\right),...,\log\left(D_{j,\left(N/2-M+i\right)}\right)\right\}$, following the derivation in Section $\ref{section general}$, we get
\begin{equation*}
\bm{\xi}=\begin{bmatrix}
\log\left(\log\left(\frac{1}{1-p}\right)\right)+\log\left(\lambda_j\right)\\
\log\left(\log2\right)+\log\left(\lambda_j\right) \\
\log\left(\log\left(\frac{1}{p}\right)\right)+\log\left(\lambda_j\right)
\end{bmatrix},
\end{equation*}
and
\begin{equation*}
\tiny
\Sigma=\begin{bmatrix}
\frac{p}{\left(1-p\right)\left(\log\left(1-p\right)\right)^2} & \frac{p}{\left(1-p\right)\log\left(1-p\right)\log\left(\frac{1}{2}\right)} & \frac{p}{\left(1-p\right)\log\left(1-p\right)\log p} \\
\frac{p}{\left(1-p\right)\log\left(1-p\right)\log\left(\frac{1}{2}\right)} & \frac{1}{\left(\log2\right)^2} & \frac{1}{\log\left(\frac{1}{2}\right)\log p} \\
\frac{p}{\left(1-p\right)\log\left(1-p\right)\log p} & \frac{1}{\log\left(\frac{1}{2}\right)\log p} & \frac{1-p}{p\left(\log p\right)^2} \\
\end{bmatrix},
\end{equation*}
thus, the asymptotic distribution of  $\hat{\mu}_{j,i}$ is normal with mean
\begin{equation*}
\small
\begin{aligned}
\mathbb{E}\left(\hat{\mu}_{j,i}\right)&=A\cdot\bm{\xi}\\
&=\frac{\alpha}{2}\log\left(\log\frac{1}{1-p}\cdot\log\frac{1}{p}\right)+\left(1-\alpha\right)\log\left(\log2\right)-\log\left(\lambda_j\right)\\
&\triangleq c\left(\alpha,p\right)-\log\left(\lambda_j\right),\\
\end{aligned}
\end{equation*}
and variance
\begin{equation*}
\begin{aligned}
\Var\left(\hat{\mu}_{j,i}\right)&=\frac{1}{N/16}A\Sigma A^T\\
&= \frac{2M}{N}\frac{\alpha^2}{4}g_1\left(p\right)+\frac{\alpha\left(1-\alpha\right)}{2}g_2\left(p\right)+\frac{\left(1-\alpha\right)^2}{\left(\log2\right)^2}\\
&\triangleq \frac{2M}{N}f\left(\alpha, p\right),\\
\end{aligned}
\end{equation*}
where
\begin{equation*}
\begin{aligned}
g_1\left(p\right)=&\frac{p}{\left(1-p\right)\left(\log\left(1-p\right)\right)^2}+\\
&\frac{1-p}{p\left(\log p\right)^2}+\frac{2p}{\left(1-p\right)\log\left(1-p\right)\log p},\\
\end{aligned}
\end{equation*}
and
\[g_2\left(p\right)=\frac{2p}{\left(1-p\right)\log\left(1-p\right)\log\frac{1}{2}}+\frac{2}{\log\frac{1}{2}\log p}.\]
\end{proof}

\noindent
{\bf Proof of Lemma~$\ref{thm1}$.}
\begin{proof}
When applying Tukey's trimean estimator $\hat{\mu}_{j,i}^T$ on $\left\{D_{j,i},D_{j,i+M},D_{j,i+2M},...,D_{j,\left(N/2-M+i\right)}\right\}$, following the derivation in Section $\ref{section Tukey}$, we have
\[\bm{\xi}_T=\begin{bmatrix}
\log\left(\frac{4}{3}\right)\lambda_j^{-1} \\
\log\left(2\right)\lambda_j^{-1} \\
\log\left(4\right)\lambda_j^{-1}\\
\end{bmatrix},\]
and
\[\Sigma_T=\begin{bmatrix}
\frac{1}{3\lambda_j^2} & \frac{1}{3\lambda_j^2} & \frac{1}{3\lambda_j^2} \\
\frac{1}{3\lambda_j^2} & \frac{1}{\lambda_j^2} & \frac{1}{\lambda_j^2} \\
\frac{1}{3\lambda_j^2} & \frac{1}{\lambda_j^2} & \frac{3}{\lambda_j^2} \\
 \end{bmatrix}_{3\times 3},\]
therefore, the asymptotic distribution of  $\hat{\mu}_{j,i}^T$ is normal with mean
\[\mathbb{E}\left(\hat{\mu}_{j,i}^T\right)=A_T\cdot\bm{\xi}_T=\frac{1}{4}\log\left(\frac{64}{3}\right)\lambda_j\triangleq c_1\lambda_j,\]
and variance
\[\Var\left(\hat{\mu}_{j,i}^T\right)=\frac{2M}{N}A_T\Sigma_T A_T^T=\frac{5M}{3N}\lambda_j^2.\]

When applying Gastwirth estimator $\hat{\mu}_{j,i}^G$ on $\left\{D_{j,i},D_{j,i+M},D_{j,i+2M},...,D_{j,\left(N/2-M+i\right)}\right\}$, following the derivation in Section $\ref{section Gastwirth}$, we have
\[\bm{\xi}_G=\begin{bmatrix}
\log\left(\frac{3}{2}\right)\lambda_j \\
\log\left(2\right)\lambda_j \\
\log\left(3\right)\lambda_j \\
\end{bmatrix},\]
and
\[\Sigma_G=\begin{bmatrix}
\frac{1}{2\lambda_j^2} & \frac{1}{2\lambda_j^2} & \frac{1}{2\lambda_j^2} \\
\frac{1}{2\lambda_j^2} & \frac{1}{\lambda_j^2} & \frac{1}{\lambda_j^2} \\
\frac{1}{2\lambda_j^2} & \frac{1}{\lambda_j^2} & \frac{2}{\lambda_j^2} \\
 \end{bmatrix},\]
therefore, the asymptotic distribution of  $\hat{\mu}_{j,i}^G$ is normal with mean
\begin{equation*}
\begin{aligned}
\mathbb{E}\left(\hat{\mu}_{j,i}^G\right)&=A_G\cdot\bm{\xi}_G \\
&=\left(0.3\times\log\left(\frac{9}{2}\right)+0.4\times\log\left(2\right)\right)\lambda_j \\
&\triangleq c_2\lambda_j,
\end{aligned}
\end{equation*}
and variance
\[\Var\left(\hat{\mu}_{j,i}^G\right)=\frac{2M}{N}A_G\Sigma_G A_G^T=\frac{1.67M}{N}\lambda_j^2.\]
\end{proof}

\noindent
{\bf Proof of Lemma~$\ref{thm3}$.}
\begin{proof}
When applying Tukey's trimean estimator $\hat{\mu}_{j,i}^T$ on $\left\{\log\left(D_{j,i}\right),\log\left(D_{j,i+M}\right),...,\log\left(D_{j,\left(N/2-M+i\right)}\right)\right\}$, following the derivation in Section $\ref{section Tukey}$, we have
\[\bm{\xi}_T=\begin{bmatrix}
\log\left(\log\left(\frac{16}{9}\right)\right)+\log\left(2\lambda_j\right) \\
\log\left(\log4\right)+\log\left(2\lambda_j\right) \\
\log\left(\log16\right)+\log\left(2\lambda_j\right)
\end{bmatrix},\]
and
\[\Sigma_T=\begin{bmatrix}
\frac{1}{3\left(\log\left(\frac{3}{4}\right)\right)^2} & \frac{1}{3\log\left(\frac{3}{4}\right)\log\left(\frac{1}{2}\right)} & \frac{1}{3\log\left(\frac{3}{4}\right)\log\left(\frac{1}{4}\right)} \\
\frac{1}{3\log\left(\frac{3}{4}\right)\log\left(\frac{1}{2}\right)} & \frac{1}{\left(\log2\right)^2} & \frac{1}{\log\left(\frac{1}{2}\right)\log\left(\frac{1}{4}\right)} \\
\frac{1}{3\log\left(\frac{3}{4}\right)\log\left(\frac{1}{4}\right)} & \frac{1}{\log\left(\frac{1}{2}\right)\log\left(\frac{1}{4}\right)} & \frac{3}{\left(\log4\right)^2} \\
 \end{bmatrix},\]
therefore, the asymptotic distribution of  $\hat{\mu}_{j,i}^T$ is normal with mean
\begin{equation*}
\begin{aligned}
\mathbb{E}\left(\hat{\mu}_{j,i}^T\right)&=A_T\cdot\bm{\xi}_T\\
&=-\left(2H+1\right)\log2j-\log2+\log\sigma^2+ \\
&\ \ \ \ \frac{1}{4}\log\left(\log\left(\frac{16}{9}\right)\cdot\log16\right)+\frac{1}{2}\log\left(\log4\right)\\
&\triangleq -\left(2H+1\right)\log2j+c_3\\
\end{aligned}
\end{equation*}
and variance
\begin{equation*}
\begin{aligned}
\Var\left(\hat{\mu}_{j,i}^T\right)&=\frac{2M}{N}A_T\Sigma_T A_T^T\\
&=\frac{M}{2N}\bigg(\frac{1}{12\left(\log\frac{3}{4}\right)^2}+\frac{1}{3\log\frac{3}{4}\log\frac{1}{2}}+\frac{1}{6\log\frac{3}{4}\log\frac{1}{4}}+ \\
&\ \ \ \ \frac{1}{\left(\log\frac{1}{2}\right)^2}+\frac{1}{\log\frac{1}{2}\log\frac{1}{4}}+\frac{3}{4\left(\log\frac{1}{4}\right)^2}\bigg)\\
&\triangleq V_1.\\
\end{aligned}
\end{equation*}

When applying Gastwirth estimator $\hat{\mu}_{j,i}^G$ on $\left\{\log\left(D_{j,i}\right),\log\left(D_{j,i+M}\right),...,\log\left(D_{j,\left(N/2-M+i\right)}\right)\right\}$, following the derivation in Section $\ref{section Gastwirth}$, we have
\[\bm{\xi}_G=\begin{bmatrix}
\log\left(\log\left(\frac{9}{4}\right)\right)+\log\left(2\lambda_j\right) \\
\log\left(\log4\right)+\log\left(2\lambda_j\right) \\
\log\left(\log9\right)+\log\left(2\lambda_j\right) \\
\end{bmatrix},\]
and
\[\Sigma_G=\begin{bmatrix}
\frac{1}{2\left(\log\frac{2}{3}\right)^2} & \frac{1}{2\log\left(\frac{2}{3}\right)\log\left(\frac{1}{2}\right)} & \frac{1}{2\log\left(\frac{1}{3}\right)\log\left(\frac{2}{3}\right)} \\
\frac{1}{2\log\left(\frac{2}{3}\right)\log\left(\frac{1}{2}\right)} & \frac{1}{\left(\log2\right)^2} & \frac{1}{\log\left(\frac{1}{2}\right)\log\left(\frac{1}{3}\right)} \\
\frac{1}{2\log\left(\frac{1}{3}\right)\log\left(\frac{2}{3}\right)} & \frac{1}{\log\left(\frac{1}{2}\right)\log\left(\frac{1}{3}\right)} & \frac{2}{\left(\log3\right)^2} \\
 \end{bmatrix},\]
therefore, the asymptotic distribution of  $\hat{\mu}_{j,i}^G$ is normal with mean
\begin{equation*}
\begin{aligned}
\mathbb{E}\left(\hat{\mu}_{j,i}^G\right)&=A_g\cdot\bm{\xi}_G\\
&=-\left(2H+1\right)\log2j-\log2+\log\sigma^2+ \\
&\ \ \ \ 0.3\times\log\left(\log\left(\frac{9}{4}\right) \right)+0.4\times\log\left(\log4\right)+ \\
&\ \ \ \ 0.3\times\log\left(\log9\right)\\
&\triangleq -\left(2H+1\right)\log2j+c_4\\
\end{aligned}
\end{equation*}
and variance
\begin{equation*}
\begin{aligned}
\Var\left(\hat{\mu}_{j,i}^G\right)&=\frac{2M}{N}A_G\Sigma_G A_G^T \\
&=\frac{2M}{N}\bigg(\frac{0.09}{2\left(\log\frac{2}{3}\right)^2}+\frac{0.12}{\log\frac{2}{3}\log\frac{1}{2}}+\frac{0.09}{\log\frac{1}{3}\log\frac{2}{3}}+ \\
&\ \ \ \ \frac{0.16}{\left(\log\frac{1}{2}\right)^2}+\frac{0.24}{\log\frac{1}{2}\log\frac{1}{3}}+\frac{0.18}{\left(\log\frac{1}{3}\right)^2}\bigg) \\
&\triangleq V_2. \\
\end{aligned}
\end{equation*}
\end{proof}

\newpage
\begin{IEEEbiographynophoto}{Chen Feng} is a Ph.D. student in the H. Milton Stewart School of Industrial and Systems Engineering at Georgia Institute of Technology.
She received her B.S. degree in Applied Mathematics from Xi'an Jiaotong University (XJTU), Xi'an, China, in 2015, and Her research interests include wavelets, big data methodology, biostatistics, clinical trials, and statistical modeling in healthcare.
\end{IEEEbiographynophoto}

\begin{IEEEbiographynophoto}{Brani Vidakovic}
is a Professor of Statistics at the School of Industrial and Systems Engineering and Department Biomedical Engineering at Georgia Institute of
Technology and Emory University.
His research focuses on biostatistics, Bayesian methodology, and statistical modeling in multiscale domains with applications in bioengineering, medicine and environmental sciences.
Vidakovic received his bachelor and masters degrees in mathematics from University of Belgrade (Serbia) in 1978,
and 1981, respectively, and Ph. D. in statistics at Purdue University (Indiana, USA) in 1992.
He was an Assistant and Associate Professor at Institute of Statistics and Decision Sciences at Duke University
form 1992 to 2000 prior to joining Georgia Institute of Technology.
Vidakovic is a Fellow of American Statistical Association and elected member of International Statistical
Institute.
\end{IEEEbiographynophoto}




\bibliography{wave1}

\begin{thebibliography}{10}
\providecommand{\url}[1]{#1}
\csname url@samestyle\endcsname
\providecommand{\newblock}{\relax}
\providecommand{\bibinfo}[2]{#2}
\providecommand{\BIBentrySTDinterwordspacing}{\spaceskip=0pt\relax}
\providecommand{\BIBentryALTinterwordstretchfactor}{4}
\providecommand{\BIBentryALTinterwordspacing}{\spaceskip=\fontdimen2\font plus
\BIBentryALTinterwordstretchfactor\fontdimen3\font minus
  \fontdimen4\font\relax}
\providecommand{\BIBforeignlanguage}[2]{{%
\expandafter\ifx\csname l@#1\endcsname\relax
\typeout{** WARNING: IEEEtran.bst: No hyphenation pattern has been}%
\typeout{** loaded for the language `#1'. Using the pattern for}%
\typeout{** the default language instead.}%
\else
\language=\csname l@#1\endcsname
\fi
#2}}
\providecommand{\BIBdecl}{\relax}
\BIBdecl

\bibitem{Moloney:2006:LDC:1183456.1183460}
\BIBentryALTinterwordspacing
K.~P. Moloney, J.~A. Jacko, B.~Vidakovic, F.~Sainfort, V.~K. Leonard, and
  B.~Shi, ``Leveraging data complexity: Pupillary behavior of older adults with
  visual impairment during hci,'' \emph{ACM Trans. Comput.-Hum. Interact.},
  vol.~13, no.~3, pp. 376--402, Sep. 2006. [Online]. Available:
  \url{http://doi.acm.org/10.1145/1183456.1183460}
\BIBentrySTDinterwordspacing

\bibitem{2017arXiv170304180K}
M.~{Kang} and B.~{Vidakovic}, ``{MEDL and MEDLA: Methods for Assessment of
  Scaling by Medians of Log-Squared Nondecimated Wavelet Coefficients},''
  \emph{ArXiv e-prints}, Mar. 2017.

\bibitem{Kolmogorov1940}
K.~AN, ``Wienersche spiralen und einige andere interessante kurven im
  hilbertschen raum,'' \emph{Acad Sci USSR (NS)}, vol.~26, pp. 115--118, 1940.

\bibitem{mandelbrot1968fractional}
B.~B. Mandelbrot and J.~W. Van~Ness, ``Fractional brownian motions, fractional
  noises and applications,'' \emph{SIAM review}, vol.~10, no.~4, pp. 422--437,
  1968.

\bibitem{abry2000wavelets}
P.~Abry, P.~Flandrin, M.~S. Taqqu, D.~Veitch \emph{et~al.}, ``Wavelets for the
  analysis, estimation and synthesis of scaling data,'' \emph{Self-similar
  network traffic and performance evaluation}, pp. 39--88, 2000.

\bibitem{abry1995wavelets}
P.~Abry, P.~Gon{\c{c}}alv{\'e}s, and P.~Flandrin, ``Wavelets, spectrum analysis
  and 1/f processes,'' \emph{LECTURE NOTES IN STATISTICS-NEW YORK-SPRINGER
  VERLAG-}, pp. 15--15, 1995.

\bibitem{abry2013scaling}
P.~Abry, P.~Goncalves, and J.~L. V{\'e}hel, \emph{Scaling, fractals and
  wavelets}.\hskip 1em plus 0.5em minus 0.4em\relax John Wiley \& Sons, 2013.

\bibitem{soltani2004estimation}
S.~Soltani, P.~Simard, and D.~Boichu, ``Estimation of the self-similarity
  parameter using the wavelet transform,'' \emph{Signal Processing}, vol.~84,
  no.~1, pp. 117--123, 2004.

\bibitem{shen2007robust}
H.~Shen, Z.~Zhu, and T.~C. Lee, ``Robust estimation of the self-similarity
  parameter in network traffic using wavelet transform,'' \emph{Signal
  Processing}, vol.~87, no.~9, pp. 2111--2124, 2007.

\bibitem{tukey1977exploratory}
J.~W. Tukey, ``Exploratory data analysis,'' 1977.

\bibitem{andrews2015robust}
D.~F. Andrews and F.~R. Hampel, \emph{Robust estimates of location: Survey and
  advances}.\hskip 1em plus 0.5em minus 0.4em\relax Princeton University Press,
  2015.

\bibitem{Gas1}
G.~J. L., ``“on robust procedures,”,'' \emph{Journal of the American
  Statistical Association}, vol.~61, p. 929, 1966.

\bibitem{Gas2}
G.~J. L. and C.~M., ``“small sample behavior of some robust linear estimates
  of locations,”,'' \emph{Journal of the American Statistical Association},
  vol.~65, p. 946, 1970.

\bibitem{Gas3}
G.~J. L. and R.~H., ``“on robust linear estimators,”,'' \emph{Annals of
  Mathematical Statistics}, vol.~40, p.~24, 1969.

\bibitem{dasgupta2008asymptotic}
A.~DasGupta, \emph{Asymptotic theory of statistics and probability}.\hskip 1em
  plus 0.5em minus 0.4em\relax Springer Science \& Business Media, 2008.

\bibitem{nason1995stationary}
G.~P. Nason and B.~W. Silverman, ``The stationary wavelet transform and some
  statistical applications,'' \emph{LECTURE NOTES IN STATISTICS-NEW
  YORK-SPRINGER VERLAG-}, pp. 281--281, 1995.

\bibitem{vidakovic2009statistical}
B.~Vidakovic, \emph{Statistical modeling by wavelets}.\hskip 1em plus 0.5em
  minus 0.4em\relax John Wiley \& Sons, 2009, vol. 503.

\bibitem{percival2006wavelet}
D.~B. Percival and A.~T. Walden, \emph{Wavelet methods for time series
  analysis}.\hskip 1em plus 0.5em minus 0.4em\relax Cambridge university press,
  2006, vol.~4.

\bibitem{vidakovic2002pollen}
D.~Pollen, ``Su i (2, f [z, 1/z]) for fa subfield of c,'' \emph{Journal of the
  American Mathematical Society}, pp. 611--624, 1990.

\end{thebibliography}
\bibliographystyle{IEEEtran}
\end{document}